\documentclass[12pt]{article}
\usepackage{graphicx}
\usepackage{cite}
\newcommand{\fnd}[2]{\frac{\textstyle #1}{\textstyle #2}}
\newcommand{\xrm}[1]{{\textstyle \mbox{\rm #1}}}
\newcommand{\bm}[1]{\mbox{\boldmath $#1$}}
\newcommand{\Real}[1]{\Re {\it e}(#1 )}
\newcommand{\Imag}[1]{\Im {\it m}(#1 )}
\newcommand{\babar}
{{\it B}$\!${\footnotesize\it A}$\!${\it B}$\!${\footnotesize\it A$\!$R}}
\def\chie{\mbox{\raisebox{0.5ex}{$\chi$}}}
\begin{document} \baselineskip .7cm
\title{Observed $D_{s}$(2317) and tentative $D$(2100--2300) \\ as the charmed
cousins of the light scalar nonet}
\author{
Eef van Beveren\\
{\normalsize\it Centro de F\'{\i}sica Te\'{o}rica}\\
{\normalsize\it Departamento de F\'{\i}sica, Universidade de Coimbra}\\
{\normalsize\it P-3004-516 Coimbra, Portugal}\\
{\small eef@teor.fis.uc.pt}\\ [.3cm]
\and
George Rupp\\
{\normalsize\it Centro de F\'{\i}sica das Interac\c{c}\~{o}es Fundamentais}\\
{\normalsize\it Instituto Superior T\'{e}cnico, Edif\'{\i}cio Ci\^{e}ncia}\\
{\normalsize\it P-1049-001 Lisboa Codex, Portugal}\\
{\small george@ajax.ist.utl.pt}\\ [.3cm]
{\small PACS number(s): 14.40.Lb, 13.25.Ft, 13.75.Lb, 12.40.Yx} \\[.3cm]
{\small hep-ph/0305035}
}
\maketitle

\begin{abstract}
The very recently observed $D^*_{sJ}(2317)^+$ meson is described as a
quasi-bound scalar $c\bar{s}$ state in a unitarized meson model, owing its
existence to the strong $^3\!P_0$ OZI-allowed coupling to the nearby $S$-wave
$DK$ threshold. By the same mechanism, a scalar $D_0^*$(2100--2300) resonance
is predicted above the $D\pi$ threshold. These scalars are the charmed
cousins of the light scalar nonet $f_0(600)$, $f_0(980)$, $K^*_0(800)$, and
$a_0(980)$, reproduced by the same model. The standard $c\bar{n}$ and
$c\bar{s}$ charmed scalars $D_{0}$ and $D_{s0}$, cousins of the scalar nonet
$f_0(1370)$, $f_0(1500)$, $K_0^*(1430)$, and $a_0(1450)$, are predicted to lie
at about 2.64 and 2.79 GeV, respectively, both with a width of some 200 MeV.
\end{abstract}

The $D^*_{sJ}(2317)^+$ (or simply $D_s(2317)$) charmed meson, just discovered
\cite{HEPEX0304021} by the \babar\ collaboration (see also
Ref.~\cite{HEPEX0305100}), is claimed
\cite{SLAC20030428} to \em ``send theorists back to their drawing boards,'' \em
in view of its low mass. If indeed the tentative $J^P=0^+$ assignment gets
confirmed, there appears to be a discrepancy with typical quark potential
models, which predict a mass of 2.48 \cite{PRD43p1679} or 2.49 GeV
\cite{PRD64p114004} for this state.
 
The aim of this Letter is to demonstrate that there is no difficulty
\cite{Ds2317} to obtain
a low-mass scalar $D_s(2317)$, with a standard $c\bar{s}$ configuration,
\em provided \em \/one takes into account its OZI-allowed coupling to the
nearby but closed $DK$ threshold at about 2.36 GeV. The strong $^3\!P_0$
coupling to this $S$-wave threshold will turn out to conjure up a
nonperturbative pole, \em not \em \/related to the confinement spectrum, which
is forced to settle down on the real energy axis due to the lack of phase
space.  Conversely, the same threshold pushes the ground-state \em confinement
\em \/pole to much higher energies, which may thus have precluded its detection
so far. By the same token and as a spin-off, we shall also predict two
additional charmed scalar mesons, i.e., with a $c\bar{n}$ ($n=u$ or $d$)
configuration, still needing experimental confirmation.

The mechanism responsible for the low-mass charmed scalar mesons is precisely
the same that produces the light scalar-meson nonet, i.e., the $f_{0}(600)$,
$f_{0}(980)$, $a_{0}(980)$ \cite{PRD66p010001}, and
$K^{\ast}_{0}(800)$ \cite{PRL89p121801}. The latter scalar mesons were
predicted with great accuracy in Ref.~\cite{ZPC30p615}, in the framework of a
unitarized quark model for all mesons. In more detail, and employing a simpler
yet less model-dependent formulation, it has been shown 
\cite{EPJC22p493,HEPPH0110156,HEPPH0201006} how unitarization leads to
structures in the scattering amplitude for $S$-wave meson-meson scattering
which are not and even \em cannot be \em  \/anticipated by the naive quark
model. Especially, from the behavior of the scattering poles near threshold
\cite{HEPPH0207022,HEPPH0304105}, we learn that, in the limit of decoupling
from the meson-meson continuum, all members of the light scalar-meson nonet
disappear into the background. On the other hand, all other mesons end up as
genuine $q\bar{q}$ states in this limit.

\begin{figure}[ht]
\normalsize
\begin{center}
\begin{picture}(273.46,293.46)(-40.00,0.00)
\put(10.66,259.94){\makebox(0,0)[bc]{\bf 2.0}}
\put(71.96,259.94){\makebox(0,0)[bc]{\bf 2.2}}
\put(133.26,259.94){\makebox(0,0)[bc]{\bf 2.4}}
\put(7.86,65.67){\makebox(0,0)[rc]{\bf -0.8}}
\put(7.86,112.85){\makebox(0,0)[rc]{\bf -0.6}}
\put(7.86,160.04){\makebox(0,0)[rc]{\bf -0.4}}
\put(7.86,207.23){\makebox(0,0)[rc]{\bf -0.2}}
\put(234.34,259.94){\makebox(0,0)[br]{\bm{\Real{E}}}}
\put(7.86,4.01){\makebox(0,0)[br]{\bm{\Imag{E}}}}
\put(27.01,63.34){\makebox(0,0){$\bullet$}}
\put(43.62,93.20){\makebox(0,0){$\bullet$}}
\put(62.23,139.46){\makebox(0,0){$\bullet$}}
\put(54.11,205.98){\makebox(0,0){$\bullet$}}
\put(26.89,236.44){\makebox(0,0){$\bullet$}}
\put(148.48,241.94){\makebox(0,0){$\bullet$}}
\put(175.39,223.92){\makebox(0,0){$\bullet$}}
\put(209.13,230.07){\makebox(0,0){$\bullet$}}
\footnotesize
\put(44.21,63.34){\makebox(0,0){\bf 0.15}}
\put(58.52,93.20){\makebox(0,0){\bf 0.2}}
\put(77.12,139.46){\makebox(0,0){\bf 0.3}}
\put(69.01,205.98){\makebox(0,0){\bf 0.5}}
\put(44.09,236.44){\makebox(0,0){\bf 0.75}}
\put(135.38,241.94){\makebox(0,0){\bf 0.3}}
\put(177.28,214.92){\makebox(0,0){\bf 0.5}}
\put(218.33,221.07){\makebox(0,0){\bf 0.75}}
\put(155,10){\makebox(0,0)[bc]{\large\bf \bm{D\pi} \bm{S} wave}}
\put(72,165){\vector(0,1){14.1}}
\put(77,168){\makebox(0,0)[lt]{\large\bf \bm{\lambda}}}
\put(147,232){\vector(1,-1){10}}
\put(152,220){\makebox(0,0)[rt]{\large\bf \bm{\lambda}}}
\normalsize
\end{picture}
\end{center}
\normalsize
\caption[]{$S$-matrix poles for $D\pi$ $S$-wave scattering
as a function of the coupling constant $\lambda$.
Threshold is at 2.009 GeV; units are in GeV.}
\label{Dn}
\end{figure}
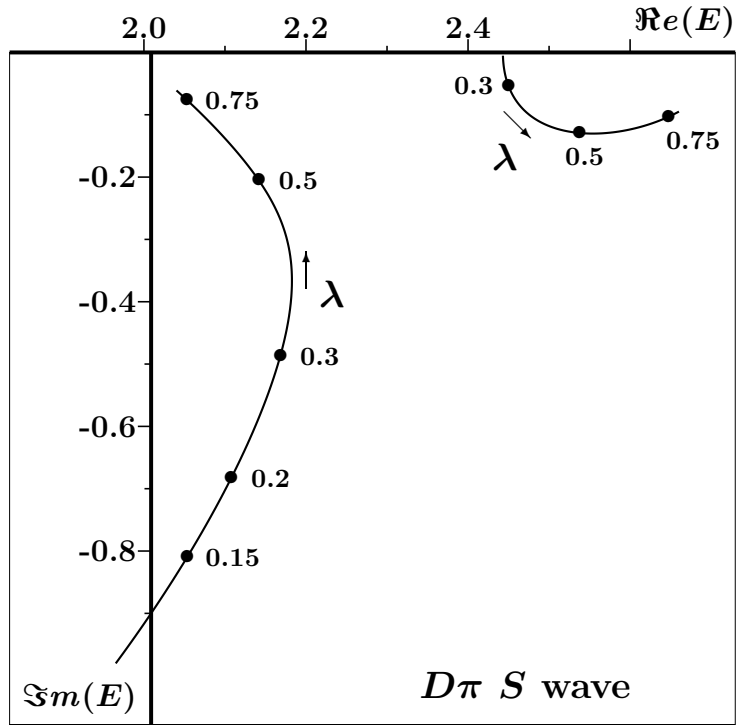

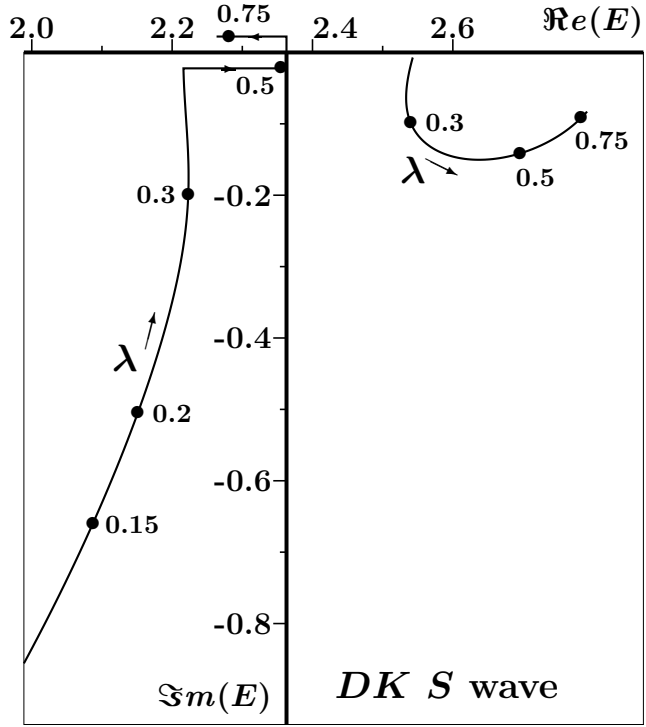
\begin{figure}[ht]
\normalsize
\begin{center}
\begin{picture}(233.46,293.46)(0.00,0.00)
\put(3.0,259.94){\makebox(0,0)[bc]{\bf 2.0}}
\put(56.07,259.94){\makebox(0,0)[bc]{\bf 2.2}}
\put(109.16,259.94){\makebox(0,0)[bl]{\bf 2.4}}
\put(162.26,259.94){\makebox(0,0)[bc]{\bf 2.6}}
\put(93.82,38.21){\makebox(0,0)[rc]{\bf -0.8}}
\put(93.82,92.26){\makebox(0,0)[rc]{\bf -0.6}}
\put(93.82,146.31){\makebox(0,0)[rc]{\bf -0.4}}
\put(93.82,200.36){\makebox(0,0)[rc]{\bf -0.2}}
\put(234.34,259.94){\makebox(0,0)[br]{\bm{\Real{E}}}}
\put(93.82,4.01){\makebox(0,0)[br]{\bm{\Imag{E}}}}
\put(75,248){\vector(1,0){5}}
\put(90,260.1){\vector(-1,0){5}}
\put(25.99,75.86){\makebox(0,0){$\bullet$}}
\put(43.04,117.81){\makebox(0,0){$\bullet$}}
\put(62.18,200.57){\makebox(0,0){$\bullet$}}
\put(97.16,248.46){\makebox(0,0){$\bullet$}}
\put(77.44,260.46){\makebox(0,0){$\bullet$}}
\put(146.30,227.76){\makebox(0,0){$\bullet$}}
\put(187.55,216.14){\makebox(0,0){$\bullet$}}
\put(210.66,229.49){\makebox(0,0){$\bullet$}}
\footnotesize
\put(41,75.86){\makebox(0,0){\bf 0.15}}
\put(56,117.81){\makebox(0,0){\bf 0.2}}
\put(50,200.57){\makebox(0,0){\bf 0.3}}
\put(95,245){\makebox(0,0)[tr]{\bf 0.5}}
\put(73,266){\makebox(0,0)[bl]{\bf 0.75}}
\put(159.28,227.76){\makebox(0,0){\bf 0.3}}
\put(193.52,207){\makebox(0,0){\bf 0.5}}
\put(218.62,221){\makebox(0,0){\bf 0.75}}
\put(160,10){\makebox(0,0)[bc]{\large\bf \bm{DK} \bm{S} wave}}
\put(46,142){\vector(1,4){3.5}}
\put(43,144){\makebox(0,0)[rt]{\large\bf \bm{\lambda}}}
\put(152,214){\vector(2,-1){12}}
\put(152,215){\makebox(0,0)[rt]{\large\bf \bm{\lambda}}}
\normalsize
\end{picture}
\end{center}
\normalsize
\caption[]{$S$-matrix poles for $DK$ $S$-wave scattering
as a function of the coupling constant $\lambda$.
Threshold is at 2.363 GeV; units are in GeV.
The trajectory of the left-hand branch partly coincides with
the real axis. For clarity, we have displaced the virtual bound states slightly
downwards, and the real bound states upwards.
Notice that for $\lambda =0.75$ (physical value) one has a real bound state
in this model.}
\label{Ds}
\end{figure}

Now, it is straightforward to apply the latter, simple model to charmed mesons:
one just has to replace one of the model's effective-quark-mass parameters
by the charmed mass. If we then take the other parameters fixed, which e.g.\
yield an excellent fit to the $S$-wave $K\pi$ phase shifts up to 1.6 GeV
\cite{EPJC22p493}, then we obtain the movement of complex-energy poles
as depicted in Figs.~\ref{Dn} and \ref{Ds}, corresponding to the $c\bar{n}$
and $c\bar{s}$ states, respectively.

Next, we interpret with some care each of the two figures, in both of which
the relevant parts of the complex $E$ plane are depicted, 
where $E$ represents the total invariant mass for elastic meson-meson
scattering in an $S$ wave.
Figure~\ref{Dn} refers to $D\pi$ and Fig.~\ref{Ds} to $DK$ scattering.
Each figure displays two trajectories, representing the positions,
as a function of the overall coupling parameter $\lambda$,
of the lowest two of an infinity of singularities
in the corresponding scattering amplitudes.

The $S$ matrix of our unitarized meson model contains, in principle,
all possible two-meson scattering channels.
In the model of Refs.~\cite{ZPC30p615,EPJC22p493,HEPPH0110156,HEPPH0201006,
HEPPH0207022,HEPPH0304105},
one single set of parameters applies to all channels, from the light flavors
to bottom.
These parameters are: four constituent quark masses, $m_{u}=m_{d}$, $m_{s}$,
$m_{c}$, and $m_{b}$, one confinement parameter, $\omega$, one overall
coupling constant, $\lambda$, and two shape parameters of the transition
potential.

When studying $D\pi$ elastic scattering, one could then select this particular
channel from a larger $S$ matrix. For the purpose of the present investigation,
we use here, as mentioned above, a simplified version of the model, 
discussed in Ref.~\cite{EPJC22p493}, which contemplates just one scattering
channel.  Nevertheless, the parameters are kept unaltered with respect to the
light scalar mesons, except for the quark masses, of course.
By comparing the results of the full model \cite{ZPC30p615}
with those of the one-channel limit of Ref.~\cite{EPJC22p493}, we verify
that the higher, closed channels do not have much influence on the general
scattering properties, but have some effect on the precise pole positions.

The two singularities studied in each of Figs.~\ref{Dn} and \ref{Ds} are the
two lowest-lying poles of the scattering amplitude.
We study their positions as a function of the overall coupling constant
$\lambda$.
The here chosen physical value of $\lambda$ equals 0.75, as in
Refs.~\cite{EPJC22p493, HEPPH0110156,HEPPH0201006,HEPPH0207022,HEPPH0304105}.
However, by just showing the respective pole positions at $\lambda =0.75$
in Figs.~\ref{Dn} and \ref{Ds}, important information on their differences
would be concealed. Moreover, the display of the pole trajectories reveals
what could happen if Nature were to choose a somewhat different value for
$\lambda$. We shall come back to this point further on.

As one observes from the two figures,
the behavior upon decoupling ($\lambda\downarrow 0$) is completely different.
Whereas the higher of the two singularities in each figure
ends up at the genuine $c\bar{n}/\bar{s}$
confinement ground state, which are at 2.44 GeV for $c\bar{n}$
and 2.55 GeV for $c\bar{s}$, respectively,  for our model parameters,
the lower poles disappear into the background, with ever increasing width.
A similar behavior has been observed for the $S$-matrix poles
of the light scalar mesons \cite{HEPPH0207022,HEPPH0304105}.

The lower pole in $D\pi$ scattering (Fig.~\ref{Dn}) does not end up below
threshold when the overall coupling $\lambda$ is increased to the
physical value $\lambda =0.75$, and settles at $E=2.03-0.075i$ GeV. Hence, in
experiment it will be observed as a structure from threshold upwards in the
partial-wave scattering cross section. For the given pole position, this
corresponds to a peak at about 2.1 GeV, with a width of 150 MeV, thus coming
out some 70 MeV higher than the real part of the pole. This shift is manifest
in our description of resonances, getting more and more sizable as the
resonance width increases. For instance, in our fit of the $S$-wave $K\pi$
phase shifts, the cross-section peak shows up more than 100 MeV above the
$K_0^*(800)$ pole \cite{EPJC22p493}. On the other hand, one should also realize
from Fig.~\ref{Dn} that a very modest decrease of $\lambda=0.75$ would give
rise not only to a larger real part of the pole position, but also of the
imaginary part, thereby amplifying the mentioned shift upwards. Such a decrease
of $\lambda$ can be justified on the basis of flavor symmetry \cite{BR03}.
Therefore, we expect a $D_0^*$ resonance somewhere in the energy interval
2.1--2.3 GeV, possibly with a width of several hundred MeV \cite{BR03}. This
may correspond to the preliminary $D_0^*(2290)$ resonance reported by the BELLE
collaboration \cite{BELLE02}, with a mass of 2.29 GeV and a width of 305 MeV.

The lower pole in $DK$ scattering (Fig.~\ref{Ds}) settles on the
real axis for $\lambda\geq0.335$.
However, for the sake of clarity we have depicted its trajectory
slightly away from the real axis.
The pole trajectory for increasing $\lambda$ ends up on the real axis
at 2.21 GeV, i.e., well below threshold.
Then it moves upwards as a virtual bound state, towards threshold,
where for $\lambda\approx 0.5$ it turns into a real bound state.
For $\lambda =0.75$ we find the pole at 2.28 GeV.
Such a behavior was already forecast in Ref.~\cite{LNP211p331}. If we were to
decrease $\lambda$ a little, again owing to flavor symmetry, we
would find \cite{BR03} a bound-state pole even closer to the experimental value
of 2.317 GeV.

The higher poles in $D\pi$ and $DK$ scattering, which stem from the 
scalar radial ground states at 2.44 GeV for $c\bar{n}$
and 2.55 GeV for $c\bar{s}$, move upwards in the second Riemann sheet.
For the physical value of the overall coupling, $\lambda =0.75$,
they constitute resonance poles. These poles come out some 200 MeV higher than
what at first sight would be expected,
not only from the naive quark model, but also from the central resonance
positions of $D_{1}(2420)$ and $D_{2}^{\ast}(2460)$ with respect to $D_{0}$,
and $D_{s1}(2536)$ and $D_{sJ}^{\ast}(2573)$ with respect to $D_{s0}$.
However, one should just compare this with the following situations,
in order to understand why our findings are not unreasonable: \\[2mm]
\indent a) \,$a_{0}(1450)$, with respect to $a_{1}(1260)$ and $a_{2}(1320)$,
\\[1mm]
\indent b) \,$f_{0}(1370)$, with respect to $f_{1}(1285)$ and $f_{2}(1270)$.
\\[2mm]
Actually, in our full unitarized model, both the $a_{0}(1450)$ and
$f_{0}(1370)$ stem from coinciding poles, connected to the model's $q\bar{q}$
scalar ground state at 1.29 GeV in the decoupling limit.

Summarizing, we have demonstrated in the foregoing how a low-mass scalar $D_s$
meson can be easily obtained by including its coupling to the most relevant
OZI-allowed two-meson channel, i.e., $DK$. However, this bound state is of a
highly nonperturbative nature, which in no way can be obtained in
single-channel quark models, no matter how sophisticated the used confinement
mechanism. In particular, we have employed a simple unitarized model, which
successfully reproduces the light scalar nonet, so as to obtain a value of 
2.28 GeV for the lowest scalar $D_s$ state, when leaving the parameters
unchanged with the exception of the quark masses. Therefore, we conclude that
the recently discovered $D_s(2317)$ is probably a scalar meson of
the latter type, being a cousin of the light scalar mesons,
rather than belonging to the scalar $q\bar{q}$ confinement spectrum, such as,
we believe, the $D_{s1}(2536)$ and $D_{sJ}^{\ast}(2573)$. Due to a similar
coupling to the $D\pi$ threshold, we predict a $D_0^*$(2100--2300) resonance,
which may have been found already \cite{BELLE02}. As a final remark
concerning the $D_s(2317)$, we should note that it becomes a bound state ---
actually a quasi-bound state owing to the, here disregarded, isospin-violating
$D_s\pi$ decay mode --- contrary to the $D_0^*$(2100--2300) and the light
scalar mesons, which are all resonances. This peculiarity is due to the high
value of the lowest OZI-allowed threshold ($DK$), while the
$D_0^*$(2100--2300) as well as \em all \em \/light scalar mesons have a
threshold involving at least one pion, which thus lie lower on a relative
scale.

Besides the latter results, we have also obtained some additional predictions
for charmed scalar mesons. For the standard charmed $q\bar{q}$ spectrum we
expect the lowest $S$-wave resonances to occur at about 2.64 GeV for $D$,
and at about 2.79 GeV for $D_{s}$, both with a width of some 200 MeV. However,
these values may be subject to a modest change when flavor-symmetry arguments
are applied as above \cite{BR03}. In any case, note
that also these two states, though being perturbative in the sense that they
can easily be linked to the confinement spectrum, turn out to suffer a drastic,
quite nonperturbative effect from unitarization.

\vspace{0.3cm}

{\bf Acknowledgments}:
We thank F.~Kleefeld for useful information on the distribution of $S$-matrix
poles in the complex $E$ and $k$ planes.
This work was partly supported by the
{\it Funda\c{c}\~{a}o para a Ci\^{e}ncia e a Tecnologia}
of the {\it Minist\'{e}rio da
Ci\^{e}ncia e da Tecnologia} \/of Portugal,
under contract number
POCTI/\-FNU/\-49555/\-2002.

\end{document}